\documentclass[a4paper,twocolumn,showpacs,superscriptaddress,floatfix,nofootinbib,hyperref]{emulateapj}
\usepackage{color,amsfonts,amsmath,amssymb,times}

\slugcomment{}

\shorttitle{On the mass radiated by coalescing black-hole binaries}
\shortauthors{Barausse, Morozova and Rezzolla}

\begin{document}


\title{On the mass radiated by coalescing black-hole binaries}


\author{E. Barausse\altaffilmark{1}}
\affil{Department of Physics, University of Guelph, Guelph, Ontario, N1G 2W1, Canada}
\altaffiltext{1}{CITA National Fellow}

\author{V. Morozova}
\affil{Max-Planck-Institut f{\"u}r Gravitationsphysik, Albert Einstein Institut, Potsdam, Germany}

\and

\author{L. Rezzolla}
\affil{Max-Planck-Institut f{\"u}r Gravitationsphysik, Albert Einstein Institut, Potsdam, Germany}
\affil{Department of Physics and Astronomy, Louisiana State University, Baton~Rouge, LA, USA}

\begin{abstract}
We derive an analytic phenomenological expression that predicts the final mass
of the black-hole remnant resulting from the merger of a generic
binary system of black holes on quasi-circular orbits. Besides
recovering the correct test-particle limit for extreme mass-ratio
binaries, our formula reproduces well the results of all the
numerical-relativity simulations published so far, both when applied
at separations of a few gravitational radii, and when applied at
separations of tens of thousands of gravitational radii. These
validations make our formula a useful tool in a variety of contexts
ranging from gravitational-wave physics to cosmology. As
representative examples, we first illustrate how it can be used to
decrease the phase error of the effective-one-body waveforms during
the ringdown phase. Second, we show that, when combined with the
recently computed self-force correction to the binding energy of
nonspinning black-hole binaries, it provides an estimate of the energy
emitted during the merger and ringdown. Finally, we use it to
calculate the energy radiated in gravitational waves by massive
black-hole binaries as a function of redshift, using different models
for the seeds of the black-hole population.
\end{abstract}


\keywords{black-hole physics --- relativity --- gravitational waves
  --- galaxies}

\maketitle

\section{Introduction}

Black-hole (BH) mergers play a central role in gravitational-wave (GW)
astrophysics, because they are expected to be among the main sources
for existing and future detectors. More specifically, the LIGO/Virgo
detectors~\citep{ligo,virgo} are expected to detect mergers of
stellar-mass BHs happening within several hundred Mpc, when operating in
their advanced configurations. Similarly, future space-based detectors
such as LISA~\citep{lisa} or DECIGO~\citep{decigo} will detect mergers 
of massive BHs (MBHs) up to redshifts as high as $z\sim 10$ or beyond. 
Even intermediate-mass BHs (IMBHs), provided they exist, will be within 
reach of GW detectors, e.g. IMBH-MBH binaries will be detectable by 
LISA or DECIGO, while IMBH-IMBH binaries will be detectable with DECIGO 
or with the planned ground-based Einstein Telescope~\citep{ET1,ET2}.

Given their relevance for GW astrophysics, it is not surprising that
BH binaries have received widespread attention over the past few
years. Because a detailed understanding of the dynamics of these
systems is crucial in order to predict accurately the gravitational
waveforms, which, in turn, is necessary to detect the signal and
extract information on the physical parameters of the binaries,
numerical simulations have been performed by a number of groups for a
variety of mass ratios, BH spin magnitudes and orientations
[see~\citet{pfeiffer2012} for a recent review].

However, even today, numerical-relativity simulations are
computationally very expensive and not able to cover the full
seven-dimensional space of parameters of quasi-circular BH
binaries. Fortunately, phenomenological models have been very
successful at reproducing many aspects of the dynamics of BH binaries
as revealed by the numerical simulations. For instance, hybrid
``phenomenological waveforms''~\citep{ajith2008,phenom}, i.e.,
templates that represent phenomenological combinations of
Post-Newtonian (PN) and numerical-relativity (NR) waveforms, can
reproduce with high precision the NR waveforms for a wide range of
binary parameters. Similar results are achieved by the
effective-one-body (EOB) model, which attempts to reproduce not only
the gravitational waveforms, but also the full dynamics of BH binaries
during the inspiral, merger and ringdown phases, by resumming the PN
dynamics~\citep{BD99,DamourResummedWfms}, and more recently the self-force
dynamics~\citep{BBL12}.

Other aspects of the dynamics of BH binaries have been
phenomenologically understood by using combinations of PN theory,
symmetry arguments, as well as hints from the test-particle limit and
fits to numerical simulations. For instance, the final spin magnitude
of the BH remnant can be predicted by a number of phenomenological
formulas~\citep{rezzolla08a, rezzolla08b, tic08, BKL, kesden,
  rezzolla08c, spin_formula}, starting from the configuration of the
binary either at small separations $r\lesssim 10 M$, or at large
separations\footnote{For MBHs, the latter are roughly the separations
  at which the dynamics starts being dominated by GW emission, and
  represent therefore the separations at which these phenomenological
  formulas should work in order to be useful in cosmological
  contexts.} $r\sim 10^4 M$. These formulas also predict the
orientation of the final spin with good accuracy when applied to
small-separation binaries, while the formula of \citet{spin_formula}
is also accurate when the binary has a large separation, e.g. $r\sim
10^4 M$, in a large portion of the parameter
space~\citep{spin_formula,emanuele}. Similar phenomenological formulas
have also been proposed for the recoil imparted to the final BH
remnant from the anisotropic emission of GWs~\citep{her07b,
  koppitz2007, rezzolla08b, superkicks, gon07, kick_RIT1, lou09,
  kick_RIT3, kick_goddard1, kick_goddard2, kick_goddard3}. Because
most of the anisotropic GW emission takes place as a result of the
strongly nonlinear merger dynamics, these recoil formulas are not
predictive, as they depend on quantities that can only be derived with
full NR simulations, but they are still useful in the statistical
studies usually performed in a cosmological
context~\citep{mymodel,statistical_kick,lou12}.

The dependence of the final mass of the BH remnant on the binary's
initial parameters has also been investigated systematically in the
literature~\citep{tic08,boy08,reisswig09,kesden,lou10}\footnote{An
  initial expression for the radiated energy was also suggested
  by~\citet{faithful_templates}, but was
  restricted to nonspinning binaries and based on early
  NR calculations.}, but the knowledge of this
dependence is far less detailed. For instance, the formula
of~\citet{tic08} [who built upon previous work by \citet{boy08}] is
calibrated to reproduce NR results for comparable-mass binaries, but
does not have the correct test-particle limit and is therefore
inaccurate for binaries with small mass ratios. The formula
of~\citet{kesden}, on the contrary, has the correct test-particle
limit, but does not reproduce accurately the NR results for
comparable-mass binaries. Finally, the formula of \citet{lou10}
depends, for generic binary configurations, on quantities that can
only be calculated using full NR simulations, and is therefore only
useful in statistical studies.

We here introduce a new phenomenological formula for the final mass of
the BH remnant (Section \ref{sec:derivation}), which, by construction,
reproduces both the test-particle limit and the regime of binaries
with comparable masses and aligned or antialigned spins, which has
been extensively investigated by NR calculations. In Section
\ref{sec:comparison} we show that this novel formula reproduces
accurately all of the available NR data (even for generic spin
orientations and mass ratios), both when applied to small- and
large-separation binary configurations.  Furthermore, in Section
\ref{sec:applications}, we consider three different areas where our
formula can be useful: \textit{(i)} we show that it can help reduce
the phase error of the EOB waveforms during the ringdown;
\textit{(ii)} we combine it with the results of \citet{LBB12} for the
self-force correction to the binding energy of nonspinning BH binaries
and derive an estimate for the energy emitted during the merger and
ringdown by nonspinning binaries; \textit{(iii)} using a
semi-analytical galaxy-formation model to follow the coevolution of
MBHs and their host galaxies, we use our formula to predict the energy
emitted in GWs by MBH binaries as a function of redshift, and show
that these predictions are strongly dependent on the model for the
seeds of the MBH population at high redshifts. Our final conclusions
are drawn in Section \ref{sec:conclusions}.

Throughout this paper, geometrized units $G=c=1$ are used.

\section{The dependence of the final mass on the spins and the mass ratio}
\label{sec:derivation}

When deriving a simple algebraic formula that expresses, with a given
precision, the mass/energy radiated by a binary system of BHs, two
regimes are particularly well-understood. On the analytic side, in
fact, the test-particle limit yields predictions that are well-known
and simple to derive. On the numerical side, the
simulations of binaries with equal-masses and spins aligned or
antialigned with the orbital angular momentum are comparatively
simpler to study, and have been explored extensively over the last few
years. Hence, it is natural that any attempt to derive an improved
expression for the radiated energy should try and match both of these
regimes. This is indeed what our formula will be built to do.

Let us therefore start by considering the test-particle limit and, in
particular, a Kerr spacetime with mass $m_1$ and spin parameter $a
\equiv S_1/m_1^2$, and a particle (or small BH) with mass $m_2$ on a
equatorial circular orbit with radius $r\gg m_1$\footnote{Without loss
  of generality, we can assume that the particle moves on a prograde
  orbit (i.e. in the positive-$\phi$ direction), and let the spin of
  the Kerr BH point up ($a>0$) or down ($a<0$).}. To first
approximation (i.e., for mass ratios $q \equiv m_2/m_1 \ll 1$), the
particle will inspiral towards the BH under the effect of GW emission,
moving slowly (``adiabatically'') through a sequence of equatorial
circular orbits~\citep{KOadiabatic} until it reaches the innermost
stable circular orbit (ISCO), where it starts plunging, eventually
crossing the horizon. The energy $E_{\rm rad}$ emitted by the particle
during the inspiral from $r\gg m_1$ to the moment it merges with the
central BH can be written as
\begin{align}
\label{equatorial_in}
\frac{E_{\rm rad}}{M}&= [1-\tilde{E}^{\rm eq}_{_{\rm ISCO}}(a)]\,\nu+o(\nu)\,,\\
\tilde{E}^{\rm eq}_{_{\rm ISCO}}(a)&=
\sqrt{1 - \frac{2}{3 \tilde{r}^{\rm eq}_{_{\rm ISCO}}(a)}}\,,\\
\tilde{r}^{\rm eq}_{_{\rm ISCO}}(a)&=
3+Z_{2}-{\rm sign}(a)\sqrt{(3-Z_{1})(3+Z_{1}+2 Z_{2})}\,,\label{eq:risco_eq}\\ 
Z_{1}&=1+(1-a^2)^{1/3} \left[ (1+{a})^{1/3}+(1-{a})^{1/3} \right]\,,\\
Z_{2}&=\sqrt{ 3 {a}^{2} + Z_{1}^{2}}\,.
\label{equatorial_fin}
\end{align}
Here, $M \equiv m_1+m_2$ is the total mass, $\nu \equiv m_1 m_2/M^2$
is the symmetric mass ratio, $\tilde{E}_{_{\rm ISCO}}$ and
$\tilde{r}_{_{\rm ISCO}}$ are respectively the energy per unit mass at
the ISCO and the ISCO radius in units of $m_1$~\citep{isco}, while the
remainder, $o(\nu)$, contains the higher-order corrections to the
radiated energy\footnote{\label{foot:1}We here use the Landau symbol
  $o$, so that $f=o(g)$ indicates that $f/g \to 0$ when $g \to
  0$. Similarly, we will also use the Landau symbol ${\cal O}$, where
  instead $f={\cal O}(g)$ indicates that $f/g \to {\rm const}$ when $g
  \to 0$.}. These corrections account, for instance, for the
conservative self-force effects, which affect the ISCO position and
energy~\citep{BS09,LBB12}, but also for the deviations from
adiabaticity, which arise because of the finiteness of the mass $m_2$
and which blur the sharp transition between inspiral and
plunge~\citep{BDplunge,OTplunge,Kplunge}, and, more in general, for
the energy emitted during the plunge and
merger~\citep{berti07,faithful_templates,buonanno_cook_pretorius}.

If the particle is initially on an inclined (i.e., non-equatorial)
circular orbit, GW emission will still cause it to adiabatically
inspiral through a sequence of circular
orbits~\citep{KOadiabatic}. Also, the inclination of these orbits
relative to the equatorial plane, which can be defined
as~\citep{iota1}~\footnote{As in the equatorial case, we can consider
  only prograde orbits ($0\leq\iota\leq\pi/2$) and allow $a$ to be
  either positive or negative.}
\begin{equation}
\label{eq:iota}
\cos(\iota) \equiv \frac{L_z}{\sqrt{Q+L_z^2}}\,, 
\end{equation}
with $Q$ and $L_z$ being respectively the Carter constant and the
azimuthal angular momentum, will remain approximately constant during
the inspiral~\citep{iota1,iota2}. As in the equatorial case, the
particle plunges when it reaches the ISCO corresponding to its
inclination $\iota$. Unlike in the equatorial case, though, the radius
of the ISCO as a function of $a$ and $\iota$ can only be found
numerically. An analytical expression, however, can be derived if one
considers only the spin-orbit coupling of the particle to the Kerr BH,
i.e., if one considers small spins $a\ll1$. In that case, in fact, one
can explicitly check [using, for instance, equations (4)--(5) of
  \citet{iota2}] that the ISCO location and energy depend only on the
combination $a\cos(\iota)$, so that at ${\cal O}(a)^2$, the
generalization of
expressions~\eqref{equatorial_in}--\eqref{equatorial_fin} to inclined
orbits is given by
\begin{align}
\label{genericTPL_in}
\frac{E_{\rm rad}}{M}&= [1-\tilde{E}_{_{\rm ISCO}}(a,\iota)]\,\nu+o(\nu)\,,\\
\tilde{E}_{_{\rm ISCO}}(a,\iota)&
\approx\sqrt{1 - \frac{2}{3 \tilde{r}_{_{\rm ISCO}}(a,\iota)}}\,,\\
\label{genericTPL_fin} 
\tilde{r}_{_{\rm ISCO}}(a,\iota)&
\approx\tilde{r}^{\rm eq}_{_{\rm ISCO}}(a\cos(\iota))\,,
\end{align}
where $\tilde{r}^{\rm eq}_{_{\rm ISCO}}$ is given
by~\eqref{eq:risco_eq}.
Expressions~\eqref{genericTPL_in}--\eqref{genericTPL_fin} reduce to
equations (\ref{equatorial_in})--(\ref{equatorial_fin}) in the case of
equatorial orbits ($\iota=0$) and are therefore exact in that limit,
with the exception of the higher-order terms in $\nu$.

As mentioned above, another case in which we know accurately the total
energy emitted in GWs is given by binaries of BHs with equal masses
and spins aligned or antialigned with the orbital angular
momentum. \citet{reisswig09}, for instance, showed that the energy
emitted by these binaries during their inspiral (from infinite
separation), merger and ringdown can be well described by a polynomial
fit 
\begin{equation}
\label{fit_reisswig09_orig}
\frac{E_{\rm rad}}{M} = p_0+p_1(a_1+a_2)+p_2 (a_1+a_2)^2\,, 
\end{equation}
where $a_1$ and $a_2$ are the projections of the spin parameters along
the direction $\hat{\boldsymbol{L}}$ of the orbital angular momentum
($a_i$ is therefore respectively positive/negative when the spin is
aligned/antialigned with $\hat{\boldsymbol{L}}$), and where the
fitting coefficients were found to be~\citep{reisswig09}
$p_0=0.04826$, $p_1=0.01559$ and $p_2=0.00485$, with uncertainties on
the order of $\sim 5\%$~\citep{reisswig09}. We recall that the
coefficient $p_0$ can be interpreted as the nonspinning orbital
contribution to the energy loss (which is the largest one and $\sim
50\%$ of the largest possible mass loss, which happens for
$a_1=a_2=1$), $p_1$ can instead be interpreted as the spin-orbit
contribution (which is $\lesssim 30\%$ of the largest possible loss),
while $p_2$ can be associated to the spin-spin contribution (which is
$\lesssim 20\%$ of the largest possible loss). Although the fit
proposed by~\citet{reisswig09} predicts a (shallow) minimum for the
radiated energy $E_{\rm rad}$ at $(a_1+a_2)/2\sim -0.8$, this 
minimum is (very likely) just an artifact of the fit due to the scarce
data available at that time~\citep{reisswig09}. Having now more data
to analyze, we can enforce the monotonicity of $E_{\rm rad}$ as a
function of $a_1+a_2$ by assuming $p_2 = p_1/4$, which constrains
the minimum of $E_{\rm rad}$ to be at $(a_1+a_2)/2=-1$. Interestingly,
a fitting expression of the type
\begin{equation}
\label{fit_reisswig09}
\frac{E_{\rm rad}}{M} = p_0+p_1(a_1+a_2)+\frac{p_1}{4} (a_1+a_2)^2\,,
\end{equation} 
provides an estimate of the radiated energy which is as accurate
as the one obtained with~\eqref{fit_reisswig09_orig}. Indeed, with
fitting parameters
\begin{equation}
\label{fitting_parameters}
p_0=0.04827\pm0.00039\,,\quad p_1=0.01707\pm0.00032 \,,
\end{equation}
this expression reproduces all of the available NR
data\footnote{\label{foot:2}The NR data considered is relative to the
  following references listed in alphabetical order: \citet{bak08,
    berti07, ber08, cam06, chu2009, chu12, han08, han10, kel11, lov11,
    lov12, mar08, pol07, pol11, reisswig09}.}  for the energy emitted
by equal-mass binaries with aligned or antialigned spins, to within
$\sim 0.005 M$ (except for almost maximal spins, see below).  Such an
accuracy is comparable to the typical accuracy of the data themselves,
so we can conclude that expressions
(\ref{fit_reisswig09})--(\ref{fitting_parameters}) summarize our
complete knowledge of the GW emission from this class of binaries to
date.

We note, however, that higher-order terms in the spins may be needed
in equation \eqref{fit_reisswig09} to reproduce the data for nearly
extremal spins. In fact, the maximum value for the radiated energy
predicted by our fit, i.e., $9.95 \%$ of the total mass of the binary
at infinite separation when $a_1=a_2=1$, is significantly less than
the $10.95 \%$ found by \citet{lov12} for $a_1=a_2\approx0.97$. Such a
large value for $E_{\rm rad}$ is somewhat off the general trend shown
by the other NR data for large aligned spins. However, it is clear
that higher-order spin terms may have to be added if more numerical
data for high-spin configurations becomes available and confirm this
result.

\begin{figure}
\begin{center}
\includegraphics[width=8.5cm]{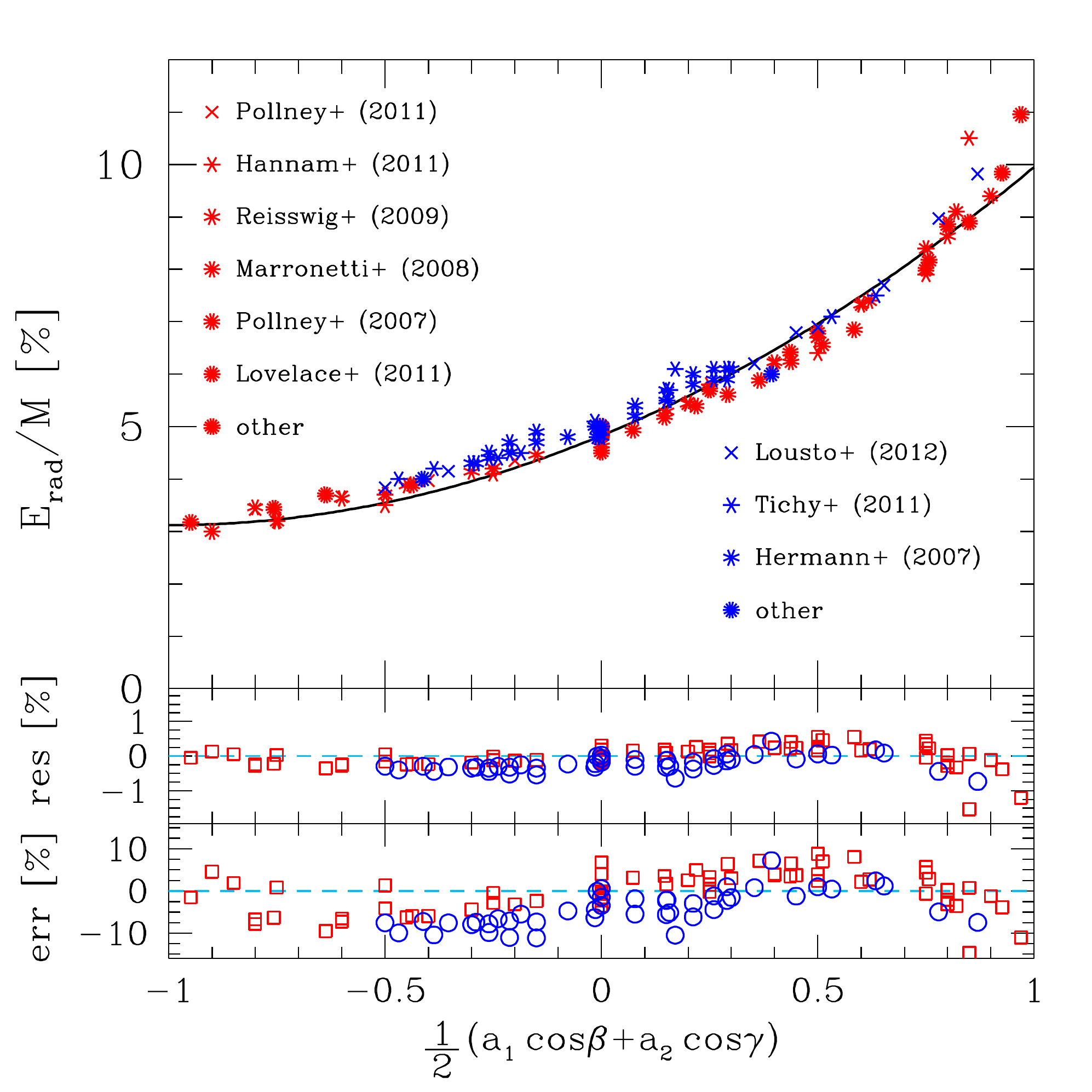}
\caption{\textit{Top panel:} Radiated energy, $E_{\rm rad}/M$, as a
  function of the total spin of the system along the orbital angular
  momentum, $|\boldsymbol{a}_1|\cos \beta +|\boldsymbol{a}_2| \cos
  \gamma$, for all published NR simulations with $q=1$, both with
  aligned/antialigned spins (in red) and for misaligned spins (in
  blue). Shown instead with a black solid line is the prediction of
  expression (\ref{fit_reisswig09_generic}) with the coefficients
  fitted from aligned/antialigned binaries. \textit{Bottom panels:}
  residuals of the NR data from the fitting expression and the
  corresponding error relative to $E_{\rm rad}/M$.}
\label{fit}
\end{center}
\end{figure}

Using therefore the knowledge of the radiated energy from the
test-particle limit and from the equal-mass aligned/antialigned
configurations, we derive an expression valid for generic binaries. 
As a first step, let us note that the PN binding
energy of an equal-mass binary of spinning BHs depends on the spins,
at 1.5 PN order, i.e., at leading order in the spins \citep{1.5PN},
only through the combination 
\begin{equation}\label{eq:LS}
\frac{\hat{\boldsymbol{L}} \cdot
(\boldsymbol{S}_1 + \boldsymbol{S}_2)}{M^2} = 
\frac{|\boldsymbol{a}_1|\cos \beta + |\boldsymbol{a}_2| \cos \gamma}{4}\,,
\end{equation}
where $|\boldsymbol{a}_1|$ and $|\boldsymbol{a}_2|$ are the spin
magnitudes, and $\beta$, $\gamma$ are the angles between the orbital
angular momentum unit vector $\hat{\boldsymbol{L}}$ and the spins of
the first and second BH, respectively. We can therefore attempt to
extend expression (\ref{fit_reisswig09}) to generic equal-mass
binaries simply by replacing $a_1+a_2$ with $|\boldsymbol{a}_1|\cos
\beta +|\boldsymbol{a}_2| \cos \gamma$, i.e.,
\begin{align}
\label{fit_reisswig09_generic}
\frac{E_{\rm rad}}{M} = p_0
&+p_1(|\boldsymbol{a}_1|\cos \beta +
|\boldsymbol{a}_2| \cos \gamma)\nonumber \\
&+\frac{p_1}{4} (|\boldsymbol{a}_1|\cos \beta +
|\boldsymbol{a}_2| \cos \gamma)^2\,.
\end{align}
As a check of this ansatz, in the top panel of Fig.~\ref{fit} we have
plotted the radiated energy, $E_{\rm rad}/M$, as a function of the
total spin along the orbital angular momentum, $|\boldsymbol{a}_1|\cos
\beta +|\boldsymbol{a}_2| \cos \gamma$, for all published NR
simulations with $q=1$, both with aligned/antialigned spins (in red;
see footnote~\ref{foot:2}) and with misaligned spins (in
blue\footnote{\label{foot:2a}The NR data for equal-mass misaligned
  binaries are relative to the following references listed in
  alphabetical order: \citet{her07a,lou12,tic07,tic08}.}). Also, we
show with a black solid line the prediction of expression
(\ref{fit_reisswig09_generic}) with the coefficients fitted from
\textit{aligned/antialigned} binaries [equation
  (\ref{fitting_parameters})]. In the bottom panels we show instead
the residuals of the NR data from the same curve and the corresponding
errors relative to $E_{\rm rad}/M$.
\begin{figure*}
\begin{center}
\includegraphics[width=8.5cm]{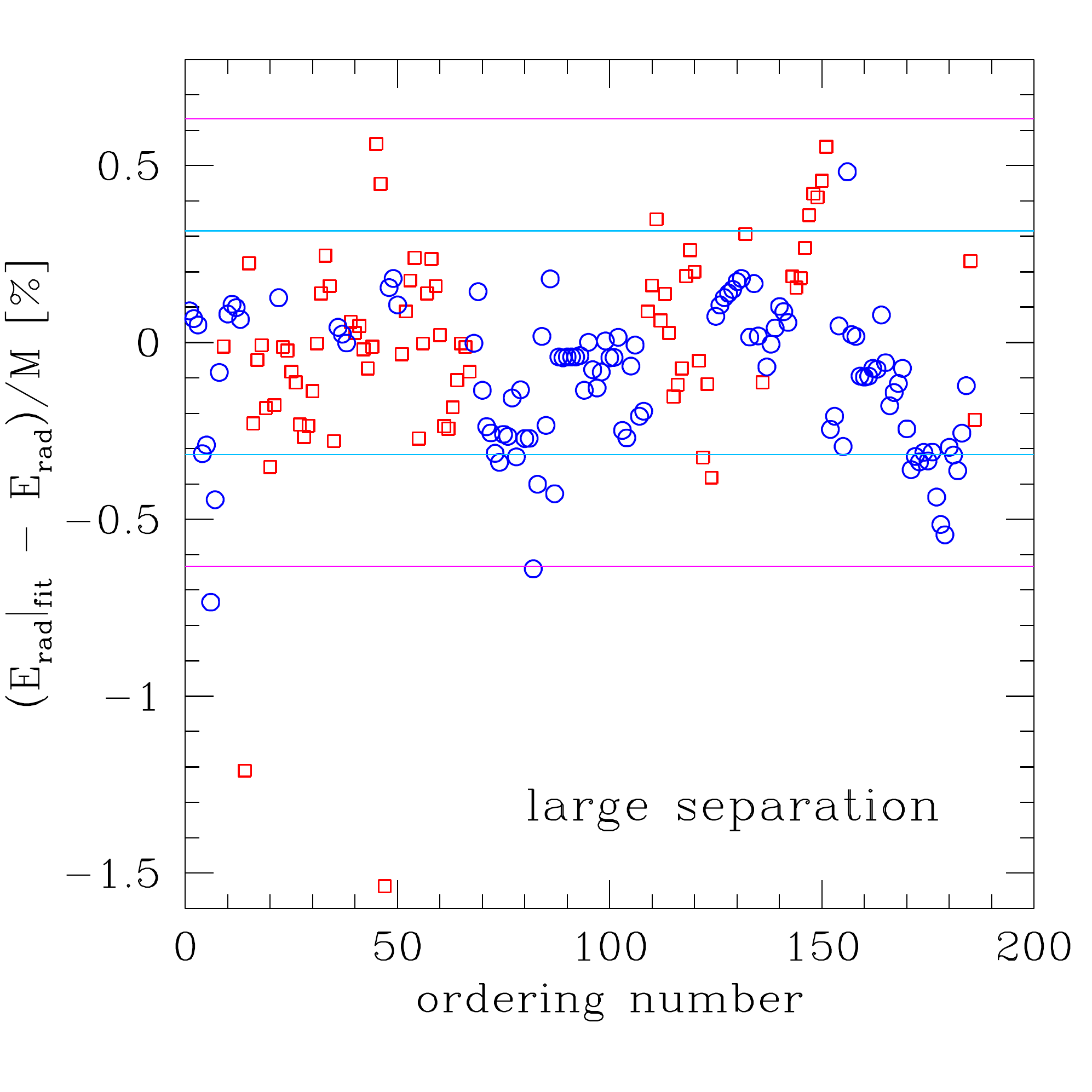}
\hskip 0.5cm
\includegraphics[width=8.5cm]{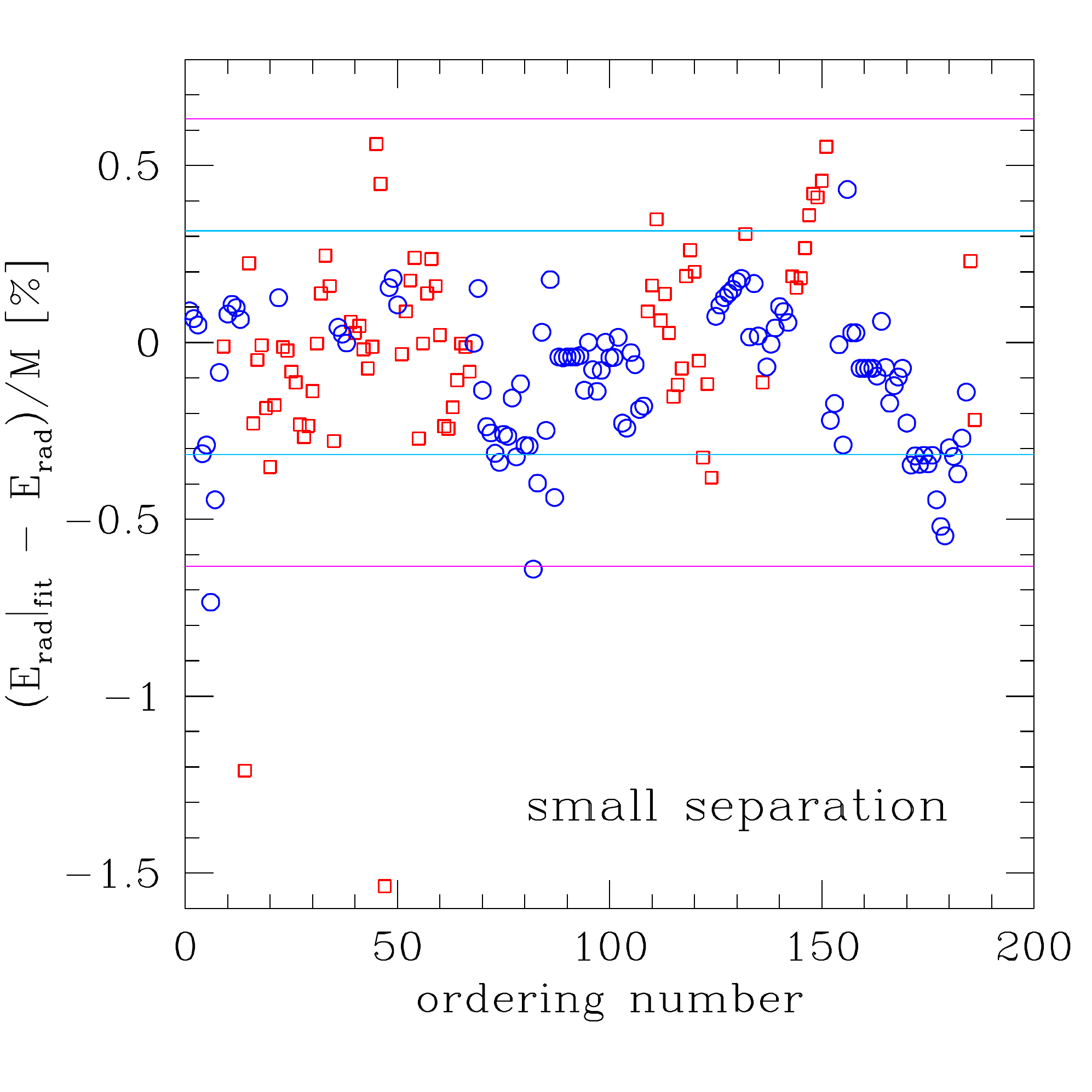}
\caption{Residuals for the fitting formula at small and large
  separations as a function of a dummy index representing the binaries
  in our dataset. Binaries with spins aligned/antialigned with the
  orbital angular momentum are plotted in red, while binaries with
  misaligned spins are shown in blue. Cyan and violet lines
  represent the $1\sigma$ and $2\sigma$ errors (estimated a
  posteriori) of the data with spins aligned/antialigned with the
  orbital angular momentum. The first $50$ points correspond to the
  simulations performed after 2010.}
\label{residuals}
\end{center}
\end{figure*}
Clearly, while future simulations that are more accurate or describe
more involved configurations may present deviations from our simple
ansatz, all published simulations for equal-mass binaries are in
reasonable agreement with equation~(\ref{fit_reisswig09_generic}),
with residuals of $\lesssim 1\%$ and errors of $\lesssim 10\%$
relative to the radiated mass. Note that these errors are comparable
with the intrinsic scatter of the different NR data.

Because in the test-particle limit the angle $\beta$ becomes the angle
between the spin $\boldsymbol{S}_1$ of the Kerr BH and the orbital
angular momentum of the particle, thus coinciding with the angle
$\iota$ defined in (\ref{eq:iota}), it is natural to rewrite equations
(\ref{genericTPL_in})--(\ref{genericTPL_fin}) as
\begin{align}
\label{eq:genericTPL_in2}
\frac{E_{\rm rad}}{M}&= 
[1-\tilde{E}_{_{\rm ISCO}}(\tilde{a})]\,\nu+o(\nu)\,,\\
\label{final_eq2}
\tilde{E}_{_{\rm ISCO}}(\tilde{a})&=
\sqrt{1 - \frac{2}{3 \tilde{r}^{\rm eq}_{_{\rm ISCO}}(\tilde{a})}}\,,
\end{align}
where we have defined
\begin{equation}
\tilde{a} \equiv 
\frac{\hat{\boldsymbol{L}}
\cdot (\boldsymbol{S}_1 + \boldsymbol{S}_2)}{M^2} 
= \frac{|\boldsymbol{a}_1|\cos \beta + q^2
|\boldsymbol{a}_2|\cos \gamma}{(1+q)^2}\,. 
\end{equation}
If we now assume that the higher-order term $o(\nu)$ in equation
(\ref{eq:genericTPL_in2}) is quadratic in $\nu$, we can determine it
by imposing that we recover the equal-mass expression
(\ref{fit_reisswig09_generic}) for $q=1$, thus obtaining the final
expression
\begin{align}
\label{final_eq1}
\frac{E_{\rm rad}}{M}=&
[1-\tilde{E}_{_{\rm ISCO}}(\tilde{a})]\,\nu \nonumber \\
& +4\,\nu^2 
[4 p_0+ 16 p_1\tilde{a} (\tilde{a}+1)+\tilde{E}_{_{\rm ISCO}}(\tilde{a})-1]\,,
\end{align}
where $\tilde{E}_{_{\rm ISCO}}(\tilde{a})$ is given
by~\eqref{final_eq2}. By construction, therefore,
expression~\eqref{final_eq1} has the correct behavior both in the
test-particle limit and for equal-mass binaries. Also, we stress that
the fitting coefficients [given by (\ref{fitting_parameters})] are
obtained using only a subset of the NR data (i.e., those for
equal-mass binaries with \textit{aligned/antialigned} spins).

\section{Comparison to data: binaries at small and large separations}
\label{sec:comparison}

In order to test the accuracy of expression
(\ref{final_eq1}), we used the data of 186
numerical simulations of inspiralling and merging BH
binaries\footnote{\label{foot:3}The data is relative to the following
  references listed in alphabetical order~\citet{bak08, ber08,
    berti07, buc12, cam06, cam09, chu2009, chu12, gon09, han08, han10,
    her07a, kel11, lou09, lou12, lov11, lov12, mar08, nak11, pol07,
    pol11, reisswig09, tic07, tic08}.}, which have reported the ratio
$M_f/M \equiv 1-E_{\rm rad}/M$ between the final mass of the BH
remnant, $M_f$, and the mass $M=m_1+m_2$ of the binary at infinite
separation. In cases where this ratio was not reported explicitly, we
have reconstructed it from the energy radiated during the numerical
simulation using PN expressions. More specifically, we calculate the
radiated energy as 
\begin{equation}
E_{\rm rad} = E^{^{\rm NR}}_{\rm rad} + |E^{^{\rm 3PN}}_{\rm bind}(\Omega_0)|\,, 
\end{equation}
where $E^{^{\rm 3PN}}_{\rm bind}(\Omega)$ is the 3PN binding energy as
function of the orbital frequency $\Omega$~\citep{buo03}, and
$\Omega_0$ is the initial orbital frequency of the simulation (either
reported explicitly or, when unavailable, reconstructed from the
initial puncture data). In addition, in those cases where the
simulation results are normalized in terms of the
Arnowitt-Deser-Misner mass $M_{\rm ADM}$, we approximate it as $M_{\rm
  ADM}=M-|E^{^{\rm 3PN}}_{\rm bind}(\Omega_0)|$.

For each binary, we apply our expression
(\ref{final_eq1}) to the initial configuration of
the numerical simulation (where the binary typically has a `` small
separation'' $r\lesssim 10 M$). However, in order to check whether our
expression predicts the final mass correctly also for widely separated
binaries, we have also integrated the initial binary back to a ``large
separation'' $r=2\times 10^4\, M$ using the quasi-circular PN
evolution equations of \citet{buo03} (which are accurate through 3.5PN
order for the adiabatic evolution of the orbital frequency, and
through 2PN order for the dynamics of the spins). For massive BHs,
this is roughly the separation at which the dynamics starts being
dominated by GW emission, and is therefore the separation at which our
expression (\ref{final_eq1}) ought to work if we
want it to be useful in cosmological contexts
[cf. \citet{spin_formula} and the discussion in \citet{mymodel}].

The results of these comparisons are summarized in
Fig.~\ref{residuals}, which shows the difference between the data and
our expression, both for small (left panel) and large separations
(right panel) as a function of a dummy index representing the ordering
of the binaries in our dataset. As can be seen the results at small
and large separations are almost indistinguishable. This does not come
as a surprise, because the projection of the total spin on the
direction of the angular momentum [equation~(\ref{eq:LS})] is
approximately conserved during the inspiral, in most of the parameter
space of quasi-circular binaries (see discussion in
\citet{spin_formula} for more details). Binaries with spins
aligned/antialigned with the orbital angular momentum are plotted in
red, while binaries with misaligned spins are shown in blue. Also
shown are cyan and violet lines representing the $1\sigma$ and
$2\sigma$ errors of the data with spins aligned/antialigned with the
orbital angular momentum, as obtained a posteriori by comparing them
to the fit~\eqref{fit_reisswig09} (cf. Fig.~\ref{fit}). Also, the
first $50$ points correspond to the simulations performed after 2010,
while others correspond to the simulations performed in
2006-2009. Although the quality of numerical simulations improved
substantially in last few years, the ``old'' data give residuals
comparable to the ``new'' ones. Furthermore, the residuals for the
binaries with spins aligned/antialigned with the orbital angular
momentum appear to be comparable with those for the binaries with
misaligned spins.

\begin{figure}[h]
\begin{center}
\vskip 0.5cm
\includegraphics[width=8.5cm]{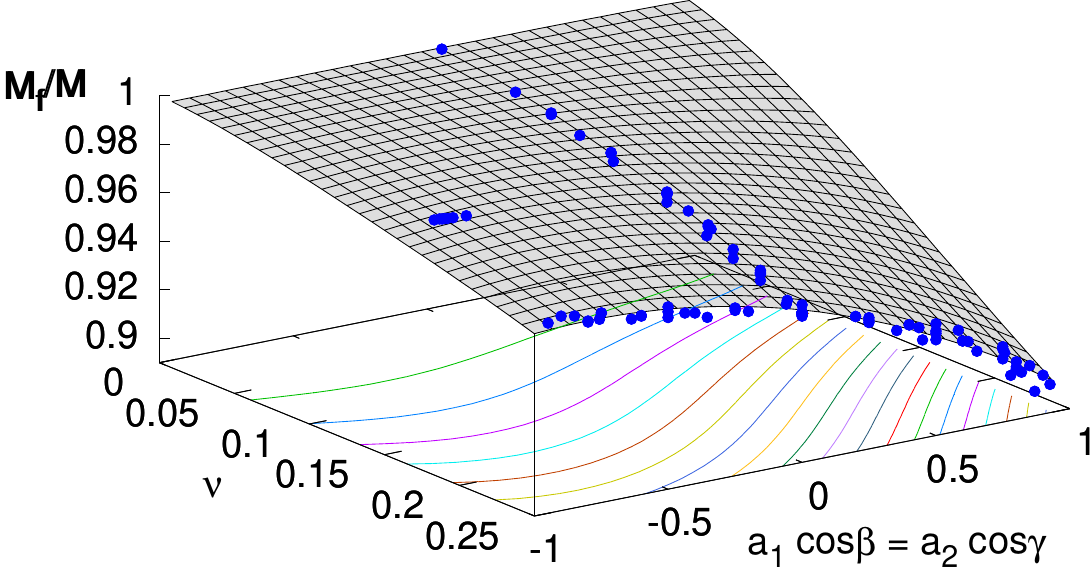}
\vskip 0.5cm
\caption{Mass of the final BH, $M_f \equiv M - E_{\rm rad}$, and
  corresponding fit for all the published binaries with $a_1 \cos
  \beta =a_2 \cos \gamma$. Note the simple functional dependence of
  the $E_{\rm rad}$, whose behaviour can be well captured with
  low-order polynomials.} \label{fit_eq_spins}
\end{center}
\end{figure}

We stress that while our expression is in reasonable agreement with
all the published data, both at large and small separations, there are
still large gaps in the parameter space of BH binaries that prevent us
from testing our approach more thoroughly. This is best seen in
Fig.~\ref{fit_eq_spins}, where we plot the final mass of the remnant for all the
published data for binaries with $a_1 \cos \beta =a_2 \cos \gamma$
(blue circles), as well as the predictions of our expression when
applied to the ``small-separation'' initial data of the simulations
(meshed surface). Clearly, spinning binaries with unequal mass ratios are
essentially absent, and simulations for such binaries will provide a
very significant check of our expression
(\ref{final_eq1}). Nevertheless, the simple
functional dependence shown by the available data, whose behaviour can
be well captured with low-order polynomials (with the possible
exception, as we already stressed, of almost maximally spinning
configurations), is quite remarkable.

The graphical representation of the data in Fig.~\ref{fit_eq_spins}
also allows to reinforce a remark already made by~\citet{reisswig09},
namely, that the largest radiated energy, $E_{\rm rad}(a=1)/M=9.95\%$,
is lost by binaries with equal-mass and maximally spinning BHs with
spins aligned with the orbital angular momentum. Hence, BH binaries on
quasi-circular orbits are among the most efficient sources of energy
in the Universe. Note, however, that equal-mass binaries are not
always the systems that lose the largest amount of energy. Indeed,
unequal-mass systems with sufficiently large spins aligned with the
angular momentum can lead to emissions larger than those from
equal-mass binaries but with large antialigned spins.  For instance, a
binary with $\nu=0.15$ and $a_1 = a_2 = 1$ will radiate more than a
binary with $\nu=0.25$ and $a_1 =- a_2$.

Notwithstanding the limited coverage of the parameter space, we can
note that our approach substantially improves upon earlier formulas
for the final mass. For instance, \citet{tic08} [building on the work
  of \citet{boy08}] suggested a formula linear in the symmetric mass
ratio $\nu$, but the coefficients needed to fit NR results are such
that the test-particle limit
(\ref{equatorial_in})--(\ref{genericTPL_fin}) is not recovered. As
mentioned earlier, because we recover the test-particle limit exactly,
our expression reproduces the published data more accurately,
especially for small mass-ratio configurations (cf. the discussion on
the effective-one-body model in the next section). Another example is
given by the formula of \citet{lou10}, which has the correct
test-particle limit but depends, for generic configurations, on
parameters that describe the binary's plunge and merger and which can
only be calculated with numerical simulations. Our algebraic formula,
instead, allows one to calculate the final mass with reasonable
accuracy, using only information on the initial binary configuration,
at any separation.

\section{Applications of the new formula}
\label{sec:applications}

In the following Sections we discuss three different examples of how
our new expression for the energy radiated during the inspiral,
merger and ringdown of two BHs can be used in contexts that range from GW
physics to cosmology.

\subsection{Merger-ringdown energy}

We can combine our expression (\ref{final_eq1})
for the total radiated energy with the recent results of \citet{LBB12}
for the binding energy of a binary of nonspinning BHs at
next-to-leading order in the mass ratio, and obtain an expression for
the energy emitted in the merger and ringdown phases of nonspinning BH
binaries. More specifically, for these binaries \citet{LBB12} found
the total energy (i.e., the binary's mass $M$ at infinite
separation, plus the binding energy) to be
\begin{equation}\label{binding_energy}
E(x)=M\left[1+\biggl( \frac{1-2x}{\sqrt{1-3x}} - 1 \biggr)\,\nu +
  \nu^2 \, E_{\rm SF}(x)\right]+{\cal O}(\nu)^3
\end{equation}
where $x \equiv (M\, \Omega)^{2/3}$ and $\Omega$ is the orbital
frequency. The self-force contribution reads
\begin{equation}\label{SFcontribution}
 E_{\rm SF}(x) = \frac{z_{\rm SF}(x)}{2} - \frac{x z'_{\rm SF}(x)}{3} 
 - 1  + \sqrt{1-3x} + \frac{x(7-24x)}{6(1-3x)^{3/2}}\,, 
\end{equation}
where $z_{\rm SF}$ is given by the fitting function 
\begin{equation}
z_{\rm SF}(x) = \frac{2 x \, (1 + a_1 x + a_2 x^2)}
{1 + a_3 x + a_4 x^2 + a_5 x^3}\,, 
\end{equation}
which is accurate to within $10^{-5}$ with $a_1 = -2.18522$, $a_2 =
1.05185$, $a_3 = -2.43395$, $a_4 = 0.400665$, and $a_5 = -5.9991$, and
where we use a prime to denote derivatives with respect to $x$.  The minimum
of $E(x)$ marks the location of the ISCO [see~\citet{LBB12,MECO}] and
lies at
\begin{gather}
x_{_{\rm ISCO}} = \frac{1}{6}\left( 1 + \frac{2}{3}\,\nu \, C_\Omega 
\right) + {\cal O}(\nu^2) \,,\\
\nonumber 
\end{gather}
with $C_\Omega= 1.2510(2)$. Replacing $x_{_{\rm ISCO}}$ in equation
(\ref{binding_energy}), one obtains that the energy emitted during the
inspiral is
\begin{align}
\label{e_insp}
\frac{E_{\rm rad, insp}}{M} 
& = 1-\frac{E(x_{_{\rm ISCO}})}{M}
\nonumber \\
& =\left(1-\frac{2\sqrt{2}}{3}\right)\nu+0.037763\, \nu^2+{\cal O}(\nu)^3 
\nonumber \\
&\simeq 0.057191\,\nu+0.037763\, \nu^2+{\cal O}(\nu)^3 \,.
\end{align}
Expressing now the total radiated energy as the sum of the one lost
during the inspiral and the one lost during the plunge-merger and
ringdown, i.e.,
\begin{equation}
\label{energy_sum}
E_{\rm rad} = E_{\rm rad, insp} + E_{\rm rad, merger-rd} \,, 
\end{equation}
and using equation (\ref{final_eq1}), we obtain
an expression for the energy radiated during the plunge-merger and ringdown
of nonspinning BH binaries:
\begin{equation}
\label{merger_rd_energy}
\frac{E_{\rm rad, merger-rd}}{M} \approx 0.506\, \nu^2\,.
\end{equation}
For equal-mass binaries, this energy is almost twice the energy lost
during the whole inspiral. In expression~\eqref{merger_rd_energy} we
have neglected terms of order ${\cal O}(\nu)^3$, so in principle this
equation may not hold for comparable-mass binaries. However, the
binding energy (\ref{binding_energy}) has been found by \citet{LBB12}
to be in very good agreement with NR results for comparable masses,
and we therefore expect the same to hold for expression
(\ref{merger_rd_energy}). Indeed, \citet{berti07} have estimated that
the energy emitted by equal- and unequal-mass nonspinning BH binaries
after the 3PN ISCO is given by
\begin{equation}\label{postISCO}
\frac{E_{\rm rad, merger-rd}}{M}\approx 0.421\, \nu^2\,.
\end{equation} 
Even more strikingly, \citet{berti07,NagarTartaglia,NagarBernuzzi} showed that the energy emitted
by a particle in a Schwarzschild spacetime during its plunge-merger is given by
\begin{equation}\label{particleE}
\frac{E_{\rm plunge}}{M}\approx 0.47 \, \nu^2\,,
\end{equation} 
in reasonable agreement with (\ref{postISCO}) and
(\ref{merger_rd_energy}). As a result, at least in the nonspinning
case, one can reproduce the radiated energy predicted by our final
expression (\ref{final_eq1}) for comparable-mass
binaries, simply by summing the energy emitted by a particle during
the inspiral [equation (\ref{e_insp})] and during the plunge-merger
[equation (\ref{particleE})].

This confirms previous work showing that perturbative results, when
expressed in terms of the symmetric mass ratio $\nu$, are in good
agreement with NR results, even for comparable masses [see
  \citet{detweiler,price,berti07} for the GW fluxes,
  \citet{periastron} for the periastron precession, and \citet{LBB12}
  for the binding energy].

\subsection{Tuning of the effective-one-body model} 

We recall that the EOB~\citep{BD99} is a phenomenological model
aiming at describing the dynamics and waveforms of BH binaries, during
the inspiral, merger and ringdown phases, combining information from PN
theory, perturbative calculations and NR. While developed
initially for nonspinning BHs~\citep{BD99,DJS3PN,BBL12}, the model has
been more recently generalized to spinning
ones~\citep{damour01,DJS,BB10,BB11,nagar}. 

To accurately describe the ringdown phase, the EOB needs expressions
predicting the final mass and spin. For non-spinning BHs, these could
be estimated self-consistently within the EOB along the lines of
\citet{DamourNagarSpin}. However, a generalization of this approach to
spinning BHs, especially if precessing, is not straightforward.
Moreover, because the final BH's mass and spin are used to calculate the
frequencies and decay times of the quasi-normal modes, even small
inaccuracies in the prediction of the remnant's mass and spin introduce
considerable phase errors in the EOB waveforms.
For
instance, the EOB model of \citet{taracchini} was compared to NR
waveforms for nonspinning BHs with mass ratio $q=1/6$, using the
formula of \citet{tic08} to calculate the final mass. Because that
formula does not have the correct test-particle limit [e.g. for
  nonspinning BHs, it predicts $E_{\rm rad}/M = 0.194\, \nu+{\cal
    O}(\nu)^2$ instead of $E_{\rm rad}/M = 0.057 \, \nu+{\cal
    O}(\nu)^2$], the EOB waveforms were accumulating a phase
difference of $\sim 0.2$ rad from the NR ones during the
ringdown. With our new formula, this phase difference during the
ringdown decreases impressively to less than $0.05$ rad (Taracchini
2012, private communication).

\subsection{GW emission by MBHs} 

According to our present understanding of galaxy formation, most
galactic nuclei should host a massive BH, with mass up to $10^{10}
\,M_\odot$. Information on the masses and spins of these BHs can be
extracted from present electromagnetic observations (see
e.g. \citet{spin_evolution} for recent constraints), but much more
accurate data will be provided by future space-based GW detectors such
as LISA or DECIGO, or terrestrial ones such as the Einstein Telescope.
These detectors will be able to observe the mergers of MBHs, which
take place when their host galaxies coalesce. Semi-analytical
galaxy-formation models, such as e.g. that of \citet{mymodel}, have
been employed to understand the MBH merger history and therefore the
event rates for these detectors, suggesting hundreds of events per
year for DECIGO, from a few to hundreds of events per year for LISA,
and up to a few tens of events per year for the Einstein Telescope. A
detector-independent diagnostics of the importance of massive BH
mergers, however, is given by the energy radiated in GWs by massive BH
binaries, per unit comoving volume and as a function of cosmic
time. We have calculated this quantity using the galaxy-formation
model of \citet{mymodel}, our new expression
(\ref{final_eq1}), and two competing models for the
seeds of the massive BHs at high redshift -- namely a ``light-seed''
scenario in which the seeds have mass $M_{\rm seed}\sim 100\, M_\odot$
at $z=15-20$, and a ``heavy-seed'' model in which the seeds form a
$z\sim 15$ with mass $M_{\rm seed}\sim 10^5\, M_\odot$ (see
\citet{mymodel} and references therein for more details).

\begin{figure}
\begin{center}
\includegraphics[width=8.5cm]{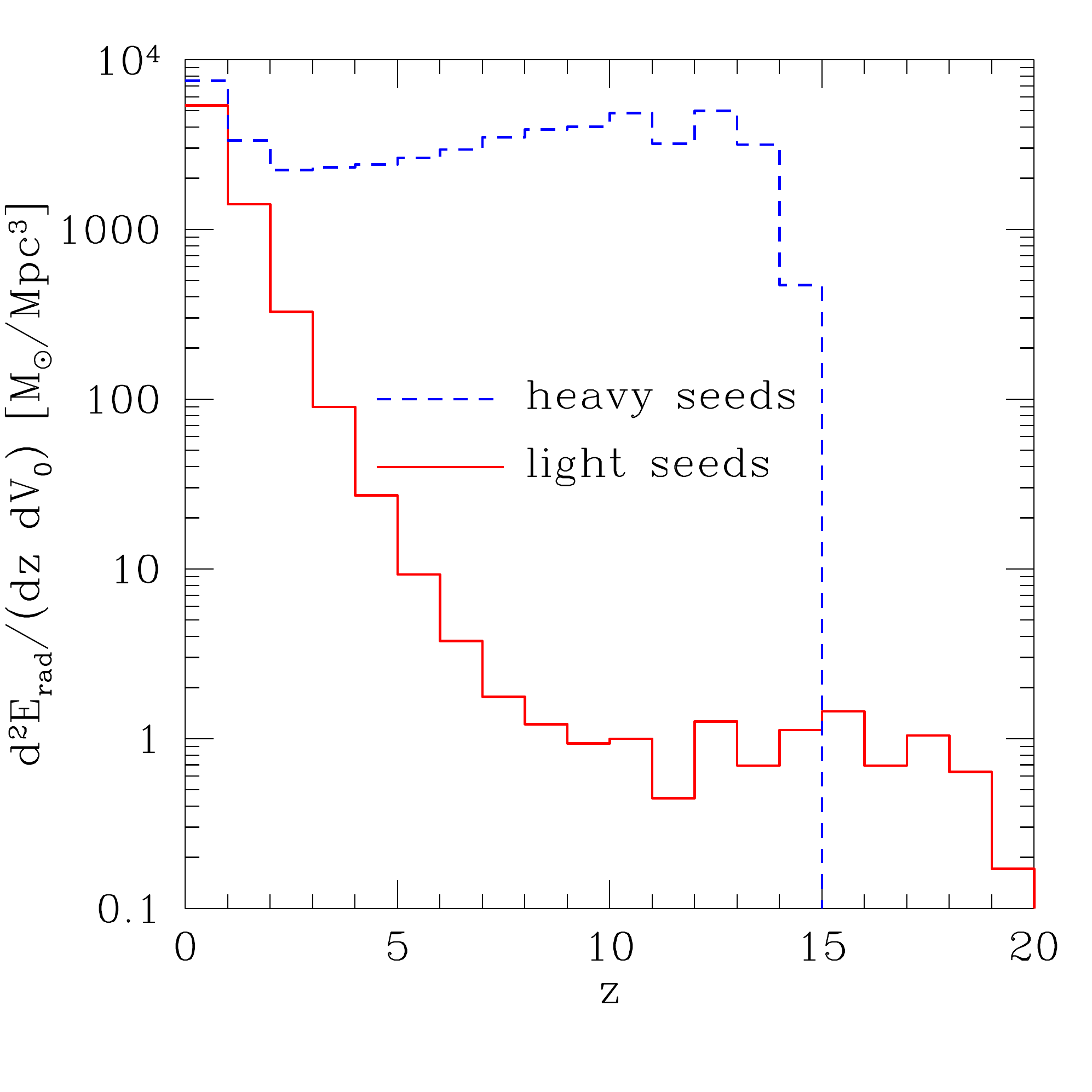}
\caption{The energy emitted by massive BH mergers per unit redshift
  and unit comoving volume, as a function of redshift. The two lines
  refer either to the ``light-seed'' scenario (red solid curve) or to
  the ``heavy-seed'' scenario (blue dashed
  line).} \label{radiated_energy_vs_z}
\end{center}
\end{figure}

The results for this quantity are shown in
Fig.~\ref{radiated_energy_vs_z}. Clearly, the heavy-seed scenario
predicts much stronger GW emission at redshifts $z\gtrsim 3$, which is
not surprising since mergers happen initially between BHs with masses
$\sim M_{\rm seed}$, and the radiated energy scales with the total
mass of the BH binaries. At $z\sim 0$, instead, the two models yield
very similar results, because both reproduce the observed local
massive BH mass function~\citep{mymodel}. Given the significantly
different yields in the GW emission expected from these two scenarios
of galaxy formation, future space-borne and terrestrial
interferometers would provide important and unambiguous information on
the BH population at early redshifts [see also
  \citet{sesana,sesana2,gair1,gair2}].

Finally, we note that by integrating the results of
Fig.~\ref{radiated_energy_vs_z}, we find that the total energy density
in GWs from massive BH binaries at $z=0$ is $\rho_{\rm
  GW,mergers}\approx7.2\times10^3\, M_\odot/{\rm Mpc}^3$ in the
light-seed scenario and $\rho_{\rm GW,mergers}\approx5.1\times10^4\,
M_\odot/{\rm Mpc}^3$ in the heavy-seed scenario, corresponding to a
cosmological density parameter $\Omega_{\rm GW,mergers} \equiv
\rho_{\rm GW,mergers}/\rho_{\rm crit}\approx5.3\times10^{-8}$
(light-seed scenario) or $\Omega_{\rm
  GW,mergers}\approx3.7\times10^{-7}$ (heavy-seed scenario).

\section{Conclusions}
\label{sec:conclusions}

We have presented a novel algebraic formula to measure the energy
radiated by coalescing binary systems of BHs with generic spin
magnitudes and orientations and arbitrary mass ratios. Our expression
uses information on the binary configuration at an arbitrary
separation and reproduces correctly the two regimes in which the
radiated energy is known best, namely, the test-particle limit (which
is known analytically) and the comparable-mass case (which has been
extensively investigated with NR simulations over the last few
years). Because it smoothly interpolates these two regimes, we expect
our formula to work reasonably well also for intermediate mass
ratios. Indeed, we have verified that it reproduces the results of all
the NR simulations published so far in the literature, including those
with unequal masses, to within an error which is comparable to the
typical errors of the simulations. In addition, we have checked that
our formula works equally well when applied to binaries starting at
small separations (i.e., $r\lesssim 10M$) and at large separations
(i.e., $r\sim 10^4 M$), thus opening up the possibility of using our
expression also in cosmological contexts.

The algebraic nature of our expression makes it a useful tool in a
variety of contexts that range from GW physics to cosmology. As
representative examples we have discussed three different
applications, namely: \textit{(i)} we have shown that, when combined
with the results of \citet{LBB12} for the self-force correction to the
binding energy of nonspinning BH binaries, the new formula provides an
estimate for the energy emitted during the merger and ringdown, and
that this estimate confirms the conjecture that the results of
perturbative calculations may be successfully extrapolated to
comparable-mass binaries when expressed in terms of the symmetric mass
ratio $\nu$; \textit{(ii)} we have shown that the new formula can help
reduce the phase error of the EOB waveforms during the ringdown;
\textit{(iii)} using a semi-analytical galaxy-formation model to
follow the coevolution of MBHs and their host galaxies, we have used
our formula to predict the energy emitted in GWs by MBH binaries as a
function of redshift, and found that these predictions strongly depend
on the scenario adopted for the MBH seeds at high redshifts, thus
making GW emission a powerful cosmological diagnostic. 

Additional uses
of the new formula can be easily considered and a particularly
relevant one is the impact of the mass loss on the accretion disk
surrounding the MBH binary. The dynamics of the disk, in fact, can
change considerably as a result of the very rapid change in the
gravitational mass of the system, with the formation of large shocks,
which are potentially detectable via their electromagnetic
emission~\citep{Oneill2009,Rossi2010,Zanotti2010}.

As a final remark we note that because our approach exploits knowledge
derived from NR simulations, the accuracy of the final-mass formula
can be improved as additional and more precise NR simulations,
especially with highly-spinning BHs, become available.

\acknowledgments 

We thank C. Lousto for useful advice on how to extract the initial
orbital frequency from the puncture data, E. Berti and U. Sperhake for
providing the initial orbital frequency of the simulations of
\citet{berti07} and \citet{ber08} respectively, H. P. Pfeiffer for
providing the data for the final mass of the simulations of
\citet{buc12} before their paper appeared on the preprint archive, and
T. Chu for providing the final mass of an unpublished simulation used
in \citet{taracchini} and presented in his PhD thesis~\citep{chu12}.
Special thanks go to E. Berti for several useful discussions, to
A. Buonanno for suggesting the relevance of this problem for the EOB,
and to A. Taracchini for confirming that our formula significantly
improves the EOB phase error for small mass ratios. E.B. acknowledges
support from a CITA National Fellowship at the University of
Guelph. This work was supported in part by the DFG grant
SFB/Transregio~7.

\end{document}